\documentstyle[12pt]{article}

\newcommand{\nc}{\newcommand}
\nc{\beq}{\begin{equation}}
\nc{\eeq}{\end{equation}}
\nc{\beqa}{\begin{eqnarray}}
\nc{\eeqa}{\end{eqnarray}}

\input epsf
\newwrite\ffile\global\newcount\figno \global\figno=1

\def\writedef#1{}
\def\figin{\epsfcheck\figin}\def\figins{\epsfcheck\figins}
\def\epsfcheck{\ifx\epsfbox\UnDeFiNeD
\message{(NO epsf.tex, FIGURES WILL BE IGNORED)}
\gdef\figin##1{\vskip2in}\gdef\figins##1{\hskip.5in}
\else\message{(FIGURES WILL BE INCLUDED)}%
\gdef\figin##1{##1}\gdef\figins##1{##1}\fi}
\def\figinsert{}
\def\ifig#1#2#3{\xdef#1{fig.~\the\figno}
\writedef{#1\leftbracket fig.\noexpand~\the\figno}%
\figinsert\figin{\centerline{#3}}\medskip\centerline{\vbox{\baselineskip12pt
\advance\hsize by -1truein\center\footnotesize{  Fig.~\the\figno.} #2}}
\bigskip\endinsert\global\advance\figno by1}
\def\endinsert{}

\begin{document}

\title{\large{\bf Pion Breather States in QCD}}

\author{
James N. Hormuzdiar\thanks{james.hormuzdiar@yale.edu} \\
Department of Physics, \\
Yale University, New Haven CT 06520 \\ \\
Stephen D.H.~Hsu\thanks{hsu@duende.uoregon.edu} \\
Department of Physics, \\
University of Oregon, Eugene OR 97403-5203 \\ \\   }

\date{May, 1998}

\maketitle

\begin{picture}(0,0)(0,0)
\put(350,345){YCTP-P13-98}
\put(350,360){OITS-652}
\end{picture}
\vspace{-24pt}

\begin{abstract}
We describe a class of pionic breather solutions (PBS) which appear in the
chiral lagrangian description of low-energy QCD. These configurations 
are long-lived, with lifetimes greater than $10^3$ fm/c, and could arise as 
remnants of disoriented chiral condensate (DCC) formation at RHIC. 
We show that the chiral lagrangian
equations of motion for a uniformly isospin-polarized domain 
reduce to those of the sine-gordon model. Consequently, our solutions 
are directly related to the breather solutions of sine-gordon theory 
in 3+1 dimensions. We investigate the possibility of PBS formation from
multiple domains of DCC, and show that the probability of formation is
non-negligible.
\end{abstract}

\newpage

\section{Introduction}

In recent work \cite{us}, we studied the evolution of domains of disoriented
chiral condensate (DCC) \cite{Anselm88,Anselm91} using the chiral lagrangian 
as a controlled, long-wavelength description. (For recent reviews of work on 
DCC, see \cite{Rajagopal:QGP2,BK:96}. Other studies of DCC
behavior are described in \cite{Rajagopal:NPB404}--\cite{JR:PRL}.) 
Our main interest in \cite{us} was the
effect of multiple domain interactions on isospin fluctuations and the
experimental signatures of DCC. As a byproduct
of that investigation, we discovered a class of long-lived classical
pion configurations which we shall refer to as Pion Breather States (PBS) or
Pion Balls.
In this letter we report in more detail on the properties of PBS and investigate
further the possibility of their formation in heavy ion collisions. PBS formation
would likely lead to a variety of dramatic signals at colliders such as RHIC,
and provide striking evidence for coherent, long-wavelength phenomena involving
the QCD vacuum.

We also note the equivalence between the chiral lagrangian dynamics of a 
uniformly isospin-polarized domain and the sine-gordon model. 
This equivalence implies that our PBS configurations are related to 
previously known 3+1 sine-gordon breathers \cite{SGB}.

\section{Pion Dynamics}

The evolution of soft pions is described by the chiral lagrangian, the unique low-energy
effective lagrangian with the symmetries of QCD.  To order $O(p^2)$ the lagrangian is

\beq
\label{chiralL}
{\cal L}
= {F_\pi ^2 \over 4}tr(\partial^\mu \Sigma^\dagger \partial_\mu \Sigma)
+F_\pi^2 m_\pi^2 tr\Sigma^\dagger ~ + ~ F_\pi^2 m_\pi^2 tr\Sigma~,
\eeq
where
\beq
\Sigma = e^{i\pi \cdot \vec \tau /F_\pi} ~,
\eeq
$\vec \pi$ represents the three pion fields, $\vec \tau$ are the three Pauli matrices, 
and $F_\pi$ is the
pion decay constant.  An explicit chiral symmetry breaking
term has been included to give the three pion types nonzero but degenerate masses.
We neglect isospin breaking and electromagnetism in what follows.

Since higher order terms and quantum loop effects are suppressed by powers of
the spacetime derivatives, the behavior of soft semiclassical configurations
can be approximated by the classical equations of motion derived from this lagrangian.  
In \cite{us} we described a numerical technique which avoids the constrained 
equations of motion resulting from (\ref{chiralL}) 
by using the linear sigma model in the limit of large self-coupling.
The same technique will be used for simulations of  multi-pion evolution in this paper.

For pion configurations which are uniformly isospin-polarized, 
the dynamics simplifies dramatically.  Using the identity
\beq
e^{i\vec A \cdot \vec \tau} = \cos(|A|)+i {\vec A \cdot \vec \tau \over |A|}
\sin(|A|)
\eeq
equation (\ref{chiralL}) can be written

\beq
\label{chiralL1}
{\cal L} ~~=~~
{1 \over 2} ~S( | \vec \pi | ) ^2 ~
\partial^\mu\vec \pi \cdot \partial_\mu \vec \pi
~~ + ~~ 
{1 \over 2} ~
\left( 1- S( | \vec \pi | )^2 
\right)~
{( \vec \pi \cdot \partial^\mu \vec \pi )^2  \over \vec \pi^2} 
~~ + ~~  F_\pi^2 ~ m_\pi^2 \cos \left( {|\vec \pi| \over F_\pi} \right) 
\eeq
where $$S( | \vec \pi | ) ~=~ {F_\pi \over |\vec \pi| } 
\sin \left( {|\vec \pi|\over F_\pi} \right)~~.$$
It is easy to see that if any two components of the pion field
are set to zero, the equations of motion guarantee that they remain
zero. Thus a perfectly isospin-polarized domain remains so under dynamical
evolution. Without loss of generality we can consider a configuration which
is polarized in the $\pi_0$ direction, yielding

\beq
{\cal L}
= {1 \over 2} \partial^\mu\pi_0 \partial_\mu\pi_0 
~ + ~ F_\pi^2 ~ m_\pi^2\cos\left( {\pi_0 \over F_\pi} \right)~,
\eeq
which is simply the sine-gordon lagrangian. Exact breather solutions
are known for this model in 1+1 dimensions \cite{SG}, and their
3+1 dimensional counterparts have been studied numerically \cite{SGB}.
Of course, we do not expect to
find completely isospin-polarized configurations in realistic situations.
The study of multiple domain configurations require that we retain (\ref{chiralL}) 
or (\ref{chiralL1}). Below we study PBS formation from single
as well as multiple domain initial conditions.

\section{Single Domain Evolution}

We studied the evolution of single domains with spherical
as well as cylindrical symmetry. The spherical iniital conditions
were taken as
\beq
|\vec \pi (r, z)|= A~ g(r-a)~,
\eeq
where $g(x)$ is a smoothed step function given by 
\beq
g(x) = {1 / \left( 1+e^{k_1 (x)} \right) }
\eeq
or
\beq
g(x) = A~ \left( {2 \over \pi} \right)tan^{-1}(e^{-x}) ~~.
\eeq
Both cases were used, without appreciably affecting the final results.  The
value of $a$ was chosen to fix the configuration size at 10 fm, 
and the value of $k_1$ was chosen to fix
the 'skin' size at approximately 1 fm. In the cylindrical case we used the functional form
\beq
|\vec \pi (r, z)|= A~ g(|z|-b)g(r-a)~.
\eeq

\begin{figure}[htb]
\center{
\leavevmode
\epsfbox{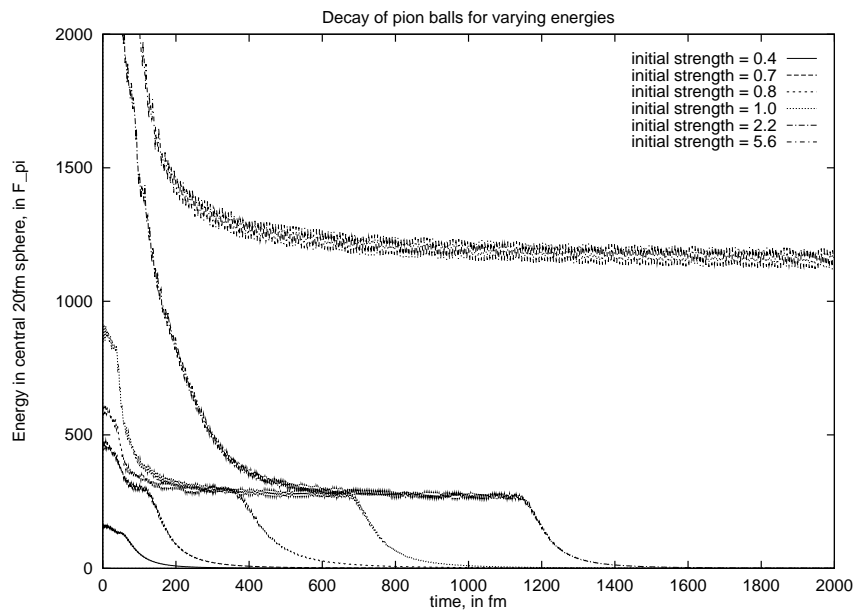}
\caption{Energy in central region vs. time} \label{enervstime}
}
\end{figure}

For sufficiently low initial strengths $s$, the pions in the initial configuration are
essentially non-interacting, and the energy disperses on a timescale of order the initial
domain size. However, at higher initial densities we observed PBS formation.
Figure~\ref{enervstime} shows a plot 
of the central energy vs. time (we define \emph{central energy}
in this paper as the energy within a fiducial volume of radius $20$ fm 
centered at the origin) 
for varying overal initial field strengths.  
As predicted, for small field strengths the central energy drops to a small
fraction of its initial value on time scales of $100$ fm, a bit larger than the initial 
configuration size.

\epsfysize=8.0 cm

\begin{figure}[htb]
\center{
\leavevmode
\epsfbox{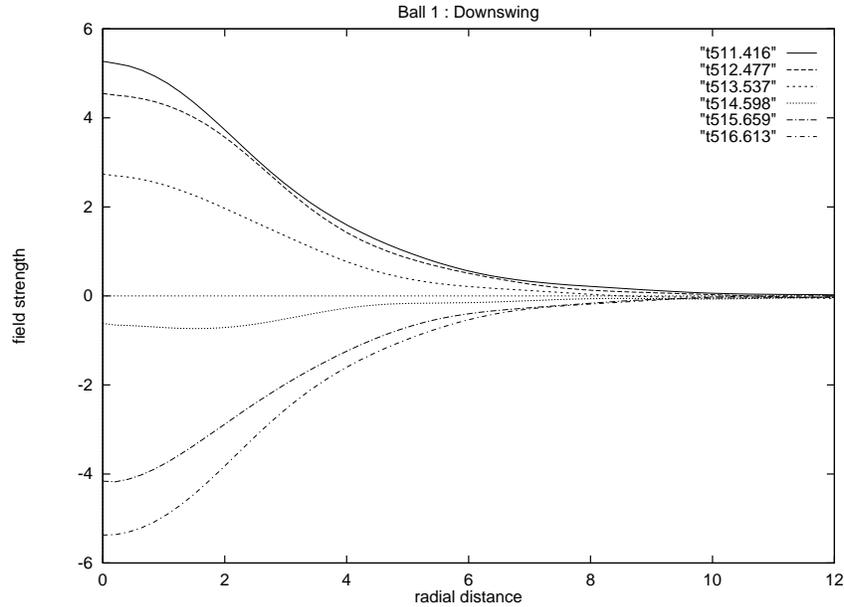}
\caption{Evolution of the lower energy PBS} \label{ball1}
}
\end{figure}

As the initial field strength is increased, PBS are formed, and their 
lifetime increases rapidly to a maximum of $1200$ fm. During their lifetime
the central energy remains almost constant at $\approx$ 26 GeV.
Inspection of the field evolution reveals an initial chaotic period during which 
packets of energy are radiated away, finally leaving a central region with nearly periodic 
oscillations of unique pattern and characteristic shape (see figure~\ref{ball1}). 
The PBS can oscillate through of order
100 cycles of period $\sim 10$ fm, amplitude $5.4 F_\pi$ and
radius $\approx 4 ~ {\rm fm}$  more or less  unchanged.  
The configuration breaks apart when small perturbations in
the oscillation pattern suddenly grow causing the shape to fall apart quickly,
again on timescales of order of $100$ fm. The decay generates approximately 200 pions.

\epsfysize=8.0 cm

\begin{figure}[htb]
\center{
\leavevmode
\epsfbox{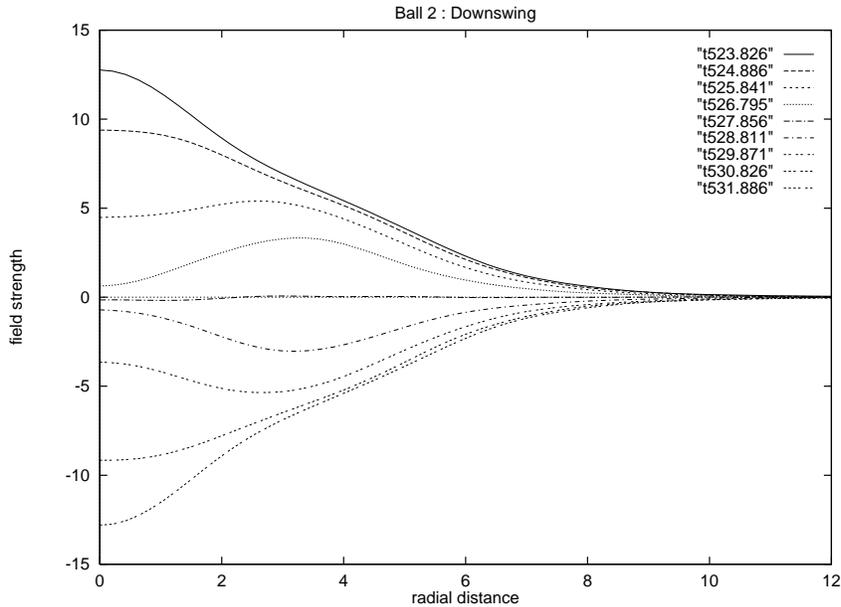}
\caption{Evolution of the higher energy PBS} \label{ball2}
}
\end{figure}

The ability of an initial configuration to form a PBS seems 
unpredictable, aside from the obvious condition that the initial energy be greater than that
of the PBS.  Lifetimes of resultant PBS are small just above threshhold and grow with increasing
energy to the maximum value. At higher energies we observed a second more energetic PBS with
energy $\approx 112$ GeV and lifetime $\approx 4000 ~ {\rm fm}$. This configuration has
a frequency of oscillation of .0619 cycles/fm, amplitude $12.9 F_\pi$ and approximate
radius of 5 fm (see figure~\ref{ball2}).  

\epsfysize=8.0 cm

The outcome of any given run is to either create one of these two PBS states, or to 
disperse the energy by around $100$ fm.  In some cases, the more energetic
PBS may decay to the other before the energy fully escapes.
We searched for novel behavior in larger and more energetic configurations, 
but none was found, suggesting the absence of 
a third type of PBS with energy less than $4000 ~ {\rm GeV}$.
Low energy cylindrically symmetric simulations were performed and yielded 
results similar to the
spherical case, with the less energetic PBS observed.

\section {Multiple Domains}

To get a better understanding of PBS formation in more realistic circumstances we
studied multiple domain initial conditions. For simplicity, we restricted ourselves
to two and three domain configurations with cylindrical symmetry. Each configuration
consists of two or three cylinders, stacked along the $\hat{z}$ axis, of radius $5.5 ~ {\rm fm}$
and height $11 ~ {\rm fm}$. The isospin orientation within each cylinder is uniform, interpolating
to the vacuum orientation at large $r$ and $|z|$.
The isospin orientation varies between domains as
\beq
\angle \vec \pi (r, z) = {2 ~ \alpha \over \pi} ~ \arctan (k_2 ~ z)~~.
\eeq
Note that isospin symmetry allows the configuration to be rotated so that the two domains fall
only in the $\pi_0-\pi_+$ plane, so only one angle need be specified.  With multiple domains this
generalizes in an obvious manner, by summing step functions with different offsets.

\begin{figure}[htb]
\center{
\leavevmode
\epsfbox{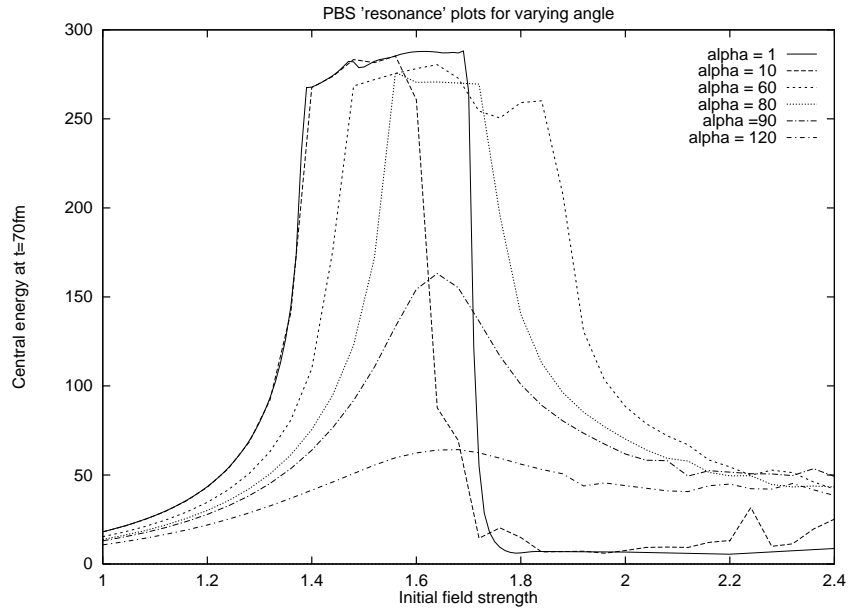}
\caption{PBS formation vs. initial field strength} \label{reson}
}
\end{figure}

We performed simulations in the two domain case, with angle $\alpha$ between the directions of the
domains in pion isospin space, and A the overall field strength.  The central energy vs.
field strength at time $70$ fm is shown in figure~\ref{reson}. $70$ fm was 
chosen as a time by which the no-PBS case will have dispersed, 
but the PBS will not have.  In the single domain case the PBS is formed sporadically but
consistently when the initial energy is above threshold. However, even a slight misalignment in
the two domain case supresses PBS formation for energies much greater than threshold.  
The remaining peak of formation exists for (roughly) $\alpha < 90^\circ$, in about 
50\% of the cases.  For three domains, we continued to find PBS formation for
a significant fraction of the cases we sampled, roughly 25\%.  
Again, an approximate rule of thumb was that
the PBS formation peak exists when the total change in the isospin orientation 
is less than $90^\circ$ over all the domains.

\epsfysize=8.0 cm

Finally, we tested the effect of incoherent contamination on the initial conditions.
To generate incoherent contamination we superposed sinusoidal field configurations
with random phases, multiplied by the same shape function of our coherent initial
condition. We found that the incoherent contamination is very destructive, 
decreasing significantly the regions of phase space that led to PBS formation.
Nevertheless, significant regions of PBS formation were found for contamination
of order 10 percent. We intend to investigate these effects in more detail in 
future work \cite{us1}.

\section{Discussion}

Our investigations indicate the following PBS properties: they are long-lived, 
can form from a robust variety of initial conditions, and could provide a
dramatic signal of non-linear QCD vacuum dynamics at relativistic heavy-ion
colliders. We intend to report in more
detail on these properties in future work \cite{us1}. However, several questions
of a theoretical nature remain unanswered. A better, analytical, understanding 
of solutions to the sine-gordon equations in 3+1 dimensions would be welcome.
In particular, we would like to know whether {\it exactly} stable solutions 
with infinite lifetimes exist. One might imagine that such solutions could
be related to previously known time-dependent, non-dissipative configurations 
\cite{Lee,Coleman}. The criteria for stability of such configurations is
generically
\beq
\label{ineq}
E < Q m~~~,
\eeq
where E is the total energy, Q a conserved charge (in our case, isospin) 
and m the mass of an asymptotic particle state. However, one can show \cite{us1} 
that with the chiral lagrangian (\ref{chiralL}) the inequality (\ref{ineq}) is
never satisfied, so any non-dissipative solutions are at best metastable.

Our attempts to find analytical solutions have so far have been unsuccessful.  
In particular, we applied the technique described in \cite{Olver} to generate 
the set of continuous symmetries for our differential equations. This
yielded that the only continuous symmetry of the spherically 
symmetric 3+1 Sine-Gordon equation is time translation, which corresponds to 
no non-trivial solution.

\bigskip
\noindent 
The authors would like to thank Alan Chodos, John Harris, Rudy Hwa, Michael Ibrahim, 
Vincent Moncrief, Dirk Rischke, Francesco Sanino, Charlie Sommerfield
and Xin-Nian Wang
for useful discussions and comments.
This work was supported in part under DOE contracts DE-AC02-ERU3075 
and DE-FG06-85ER40224.

\end{document}